\documentclass[twocolumn,showpacs,preprintnumbers,amsmath,amssymb]{revtex4}

\usepackage{graphicx}
\usepackage{dcolumn}
\usepackage{bm}

\newcommand{\betwo}{$B(E2, 0^{+}_1 \rightarrow 2^{+}_1)$ }
\newcommand{\twop}{$2^{+}$ }
\newcommand{\opoh}{$1p$-$1h$ }

\begin{document}

\preprint{Ver.1.0}

\title{Spatially extended coherence induced by pairing correlation\\
in low-frequency vibrational excitations of neutron drip line nuclei
}

\author{Masayuki Yamagami}
 \email{yamagami@riken.jp}
\affiliation{%
Heavy Ion Nuclear Physics Laboratory, RIKEN, 
Hirosawa 2-1, Wako, Saitama 351-0198, Japan
}%


\date{\today}

\begin{abstract}
Role of pairing correlation for emergence of low-frequency
vibrational excitations in neutron drip line nuclei 
is discussed paying special attention to neutrons 
with small orbital angular momentum $\ell$.    
Self-consistent pairing correlation
in the Hartree-Fock-Bogoliubov (HFB) theory
causes the change of the spatial structure 
of the quasiparticle wave functions;
"the pairing anti-halo effect" in the lower component
and "the broadening effect" in the upper component.
The resultant spatial distribution
of the two-quasiparticle states among low-$\ell$ neutrons,
"the broad localization", brings about qualitatively new aspects,
especially the large transition strength of the low-frequency
vibrational excitations in nuclei close to the neutron drip line.
By performing HFB plus 
quasiparticle random phase approximation (QRPA) calculation for
the first $2^+$ states in neutron rich Ni isotopes,
the unique role of self-consistent pairing correlation
is pointed out.
\end{abstract}

\pacs{Valid PACS appear here}
\maketitle


\section{Introduction}

Study of low-frequency vibrational excitations 
in neutron drip line nuclei is one of the most interesting subjects
in nuclear structure physics.
Vibrational excitations are microscopically
represented by coherent superposition of 
two-quasiparticle states
(or one-particle - one-hole ($1p$-$1h$) states 
in closed shell nuclei) \cite{RS80}.
In stable nuclei, the main configurations of 
the low-frequency vibrational excitations are 
two-quasiparticle states among tightly-bound single-particle states.
In neutron drip line region, by contrast, the contributing
single-particle states are tightly-bound states, 
loosely-bound states, resonant and non-resonant continuum states.
Consequently the two-quasiparticle states among them have 
rich variety of the spatial structure.
Therefore the low-frequency vibrational excitations,
as the coherent motions among such two-quasiparticle states,
may have qualitatively new aspects compared to those in stable nuclei.

We expect unique impacts of the two-quasiparticle states 
involving loosely-bound low-$\ell$ neutrons 
on the low-lying excitations.
Loosely-bound low-$\ell$ neutrons have an appreciable 
probability to be outside the core nucleus that leads to the neutron 
halo structure (see Refs \cite{Mue93,Rii94,Han95,Tan96,JR04} 
for reviews), and the coupling to the nearby continuum states 
causes the soft excitation with large transition strength 
as a non-resonant single-particle excitation \cite{SG95}.
An example is the soft dipole excitation in light halo nuclei
\cite{HJ87,Ik92,Sackett,Shimoura,Zinser,Be-Nak,Be-Pal,C-Nak,C-Dat,He}, 
typically in $^{11}$Be and $^{11}$Li, where significant $E1$
strength is observed.

On the other hand, the role of two-quasiparticle states 
among loosely-bound low-$\ell$ neutrons and 
resonant states for low-frequency vibrational excitations 
is not well understood so far.
It is still an open question whether the coherent motions
among such two-quasiparticle states can be realized
in spite of the variety of the spatial structure,
and the resultant low-lying excitations may possess 
novel features in neutron drip nuclei.

In the present study,
we emphasize the unique role of self-consistent pairing correlation
that changes the spatial structure of the quasiparticle wave functions;
"the pairing anti-halo effect" in the lower component
and "the broadening effect" in the upper component.
We show that the resultant two-quasiparticle states can cause 
the coherence among them not only in the surface region 
but also in the spatially extended region. 
Consequently the spatially extended coherence induced by
pairing correlation leads to the large collectivity of 
the low-frequency vibrational excitations.

In Sec.\ref{SEC-QPWF},
by solving the Woods-Saxon potential plus HFB pairing model,
we examine the pairing correlation in loosely bound nuclei
and the induced change of
the spatial structure of the quasiparticle wave functions.
In Sec.\ref{SEC-TQPS}
we examine the spatial structure
of two-quasiparticle states in neutron drip line nuclei.
In Sec.\ref{SEC-QRPA} we perform Skyrme-HFB plus QRPA calculation
for the first $2^{+}$ states in neutron rich Ni isotopes. 
By comparing three types of calculations; HFB plus QRPA, 
resonant BCS plus QRPA, and RPA, 
we emphasize the unique role of self-consistent pairing correlation
for realizing low-frequency vibrational excitations 
and the large collectivity in neutron drip line nuclei.
Conclusions are drawn in Sec.\ref{SEC-CONC}.


\section{Spatial structure of quasiparticle wave functions} 
\label{SEC-QPWF}


\subsection{Model} 
\label{SUBSEC-MODEL}

For our qualitative discussion, 
we solve the HFB equation in coordinate space 
\cite{Bu80,DF84,BS87,DN96} with spherical symmetry,
\begin{eqnarray}
& &\int dr'
\left( 
\begin{array}{*{20}c}
   h_{lj}(r,r') &  \Delta (r,r') \\
   \Delta(r,r') & -h_{lj} (r,r') \\
\end{array}
\right)
\left( {\begin{array}{*{20}c}
u_{lj} (E,r')  \\
v_{lj} (E,r')  \\
\end{array}} \right) \nonumber \\
& & = 
\left(
\begin{array}{*{20}c}
E+\lambda & 0 \\
0 & E-\lambda 
\end{array}
\right)
\left( 
\begin{array}{*{20}c}
u_{lj} (E,r)  \\
v_{lj} (E,r)  \\
\end{array} \right),
\label{EQ-HFB}
\end{eqnarray}
where $E$ is the quasiparticle energy, $\lambda$ is the Fermi energy,
and 
$u_{lj} (E,r)$ ($v_{lj} (E,r)$) is the upper (lower) component 
of the radial quasiparticle wave function.
The upper and lower components
are the generalization of the $u_{lj}$ and $v_{lj}$ coefficients 
in the BCS approximation.

For the mean-field hamiltonian $h_{lj}(r,r')$, 
the Woods-Saxon potential together with 
the related spin-orbit potential \cite{HM03},
\begin{eqnarray}
V_{lj}(r) = V_{WS} f(r) - V_{WS} v \left(\frac{\Lambda}{2} \right)^2 
\frac{1}{r} \frac{df(r)}{dr}\left(\vec \sigma  \cdot \vec l \right),
\label{EQ-WSPOT}
\end{eqnarray}
where
\begin{eqnarray}
f (r) = 1/\left(1 + \exp \left( \frac{r - R_{WS}}{a} \right) \right),
\label{EQ-WSFORM}
\end{eqnarray}
is adopted.
$\Lambda=\hbar/mc$ is the reduced Compton wave-length of nucleon.
The parameters are fixed to be $a=0.67$ fm and $v=32$ in accordance 
with Ref.\cite{HM03}. The strength $V_{WS}$ and the radius $R_{WS}$ 
are chosen to simulate the shell structure of nuclei 
under consideration.
For the pairing correlation,
we self-consistently derive  the pairing potential $\Delta(r,r')$
from the density dependent zero-range pairing interaction,
\begin{eqnarray}
V_{pair} (r,r') = \frac{1}{2} V_{pair} (1-P_{\sigma}) 
\left[ 1 - \frac{\rho (r)}{\rho_c} \right]
\delta (r-r'), \label{EQ-DDPI}
\end{eqnarray}
that leads to the local pairing potential,
\begin{eqnarray}
\Delta (r,r')=\Delta(r) \delta(r-r'),
\end{eqnarray}
with
\begin{eqnarray}
\Delta(r)=\frac{1}{2} V_{pair}
\left[ 1 - \frac{\rho (r)}{\rho_c} \right] 
\tilde{\rho}(r),
\label{EQ-PAIRPOT}
\end{eqnarray}
where $P_{\sigma}$ is the spin exchange operator.
The normal density $\rho(r)$ and the abnormal density $\tilde{\rho}(r)$
are defined by
\begin{eqnarray}
\rho(r)&=&\frac{1}{4\pi r^2} \sum_{lj,n} (2j+1)
\{v_{lj}(E_{lj,n},r)\}^2, \\
\tilde{\rho}(r)&=&\frac{-1}{4\pi r^2} \sum_{lj,n} (2j+1)
u_{lj}(E_{lj,n},r) v_{lj}(E_{lj,n},r).
\end{eqnarray}
To measure the pairing correlation in nuclei under consideration,
we calculate the pairing energy, 
\begin{eqnarray}
E_{pair}=\frac{1}{2} \int 
\Delta(r) \tilde{\rho}(r) 4 \pi r^2 dr.
\label{EQ-EPAIR}
\end{eqnarray}
For comparison with Ref.\cite{HM03,HS04,HM04}, 
the average pairing gap defined by
\begin{eqnarray}
\bar{\Delta} = \frac{ \int 
\Delta(r) \tilde{\rho}(r) r^2 dr}{\int 
\tilde{\rho} (r) r^2 dr},
\label{EQ-DELAV}
\end{eqnarray}
is also calculated.

\subsection{Inputs}

We examine the spatial structure of the two-quasiparticle states
of neutrons around $^{86}$Ni as the illustrative example of 
neutron drip line nuclei.
The nucleus $^{86}$Ni is at the neutron drip line
within Hartree-Fock (HF) calculation with Skyrme SLy4 force \cite{CB98},
and has the loosely-bound $3s_{1/2}$ state as the Fermi level
(see Fig.\ref{FIG_SPE}).
The Woods-Saxon potential with $V_{WS}=-41.4$ MeV and $R_{WS}=5.5$ fm
simulates the neutron shell structure, and
the obtained single-particle states around the Fermi level are 
$1g_{9/2}$ ($\varepsilon_h=-4.93$ MeV), 
$2d_{5/2}$ ($\varepsilon_h=-1.77$ MeV), 
$3s_{1/2}$ ($\varepsilon_h=-0.50$ MeV),
resonant $d_{3/2}$ ($\varepsilon_p\approx 0.14$ MeV), and
resonant $g_{7/2}$ ($\varepsilon_p\approx 0.98$ MeV).
The index $p$ ($h$) represents all necessary quantum numbers
to specify the particle (hole) state.
In the present study, the box boundary condition 
with a box size $R_{box}=75$ fm is imposed.
We elaborately examine the applicability of the box boundary
condition for describing pairing correlations in loosely bound
nuclei in the following subsections.

We solve Eq.(\ref{EQ-HFB}) only for neutrons, and 
$\rho(r)$ ($\tilde{\rho}(r)$) represents the neutron normal (abnormal)
density in the present section.
The pairing strength is fixed to be $V_{pair}=-680$ MeV fm$^{-3}$ 
with cut-off energy $E_{cut}=50$ MeV.
The parameter $\rho_c$ is taken to be
0.08 fm$^{-3}$ ($\approx \rho(0)$) 
for the surface-type pairing field.
The obtained average pairing gap is 
$\bar{\Delta}\approx 1.3 (\approx 12/\sqrt{86}$) MeV.


\subsection{Pairing correlation in loosely-bound nuclei}

\begin{figure}[tb]
\center
\includegraphics[scale=0.72]{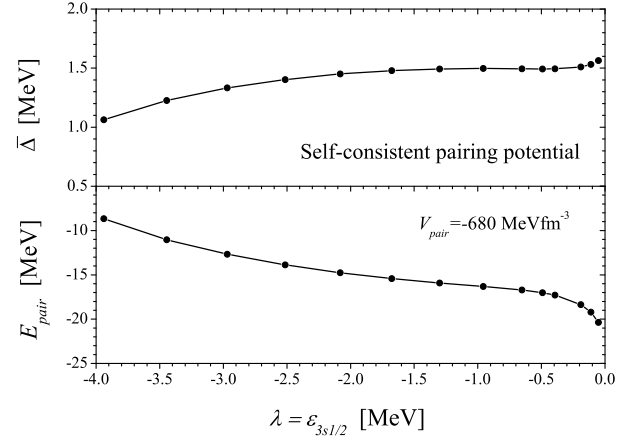}
\vspace{-8mm}
  \caption{The average pairing gap $\bar{\Delta}$ 
and the pairing energy $E_{pair}$ obtained by the Woods-Saxon
potential plus HFB pairing model are plotted as a function of 
$\lambda$$=$$\varepsilon_{3s_{1/2}}$.
The density-dependent pairing force 
with $V_{pair}$$=$$-680$ MeV fm$^{-3}$ is used.
The box size is fixed to be 75 fm.
}
\label{FIG_Epair}
\end{figure}   

\begin{figure}[tb]
\center
\includegraphics[scale=0.8]{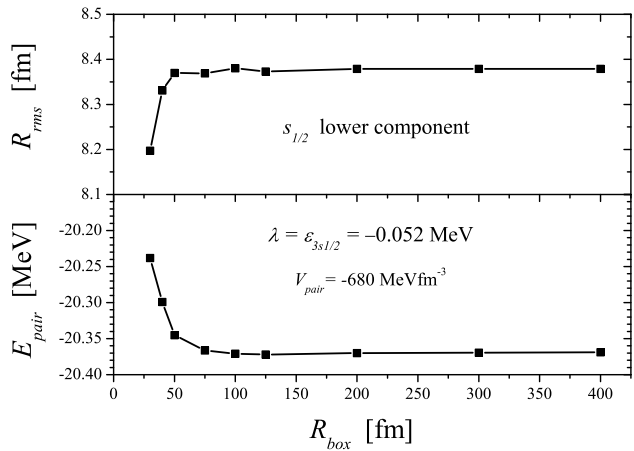}
\vspace{-8mm}
  \caption{The box size dependence of the average rms radius 
of the lower component of the $s_{1/2}$ quasiparticle wave function 
and the pairing energy obtained by the Woods-Saxon potential
plus HFB pairing model
with the condition of $\lambda=\varepsilon_{3s_{1/2}}=-0.052$ MeV. 
}
\label{FIG_Epair-cnv}
\end{figure}   

\begin{figure}[tb]
\center
\includegraphics[scale=0.8]{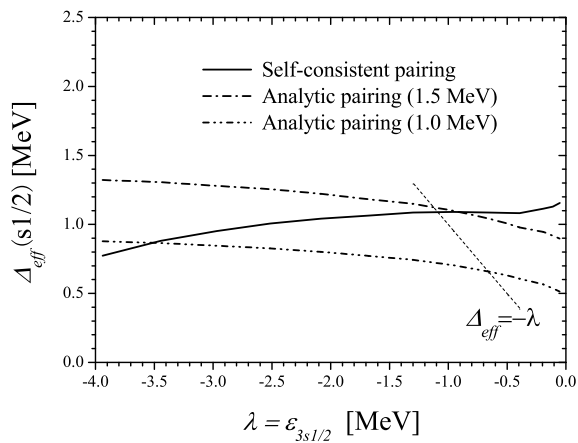}
\vspace{-8mm}
  \caption{The average effective pairing gap of the $s_{1/2}$ state 
as a function of $\lambda=\varepsilon_{3s_{1/2}}$.  
The results of the Woods-Saxon potential plus HFB pairing model
with the self-consistent pairing potential
and the analytic pairing potentials of 
$\bar{\Delta}=1.0$ and $1.5$ MeV are compared.}
\label{FIG_DELeff}
\end{figure}   

\begin{figure}[tb]
\center
\includegraphics[scale=0.75]{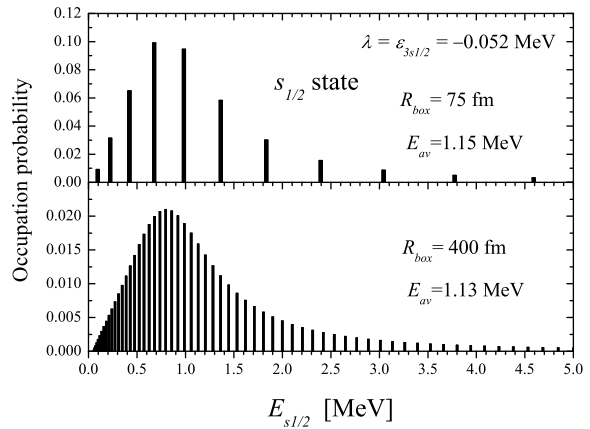}
\vspace{-8mm}
  \caption{The occupation probability of the $s_{1/2}$ state with
the condition of $\lambda$$=$$\varepsilon_{3s_{1/2}}$$=$$-0.052$ MeV 
obtained by the Woods-Saxon potential plus HFB pairing model.
The results obtained with $R_{box}=75$ and 400 fm are compared.
}
\label{FIG_s-ocp}
\end{figure}

\begin{figure}[tb]
\center
\includegraphics[scale=0.8]{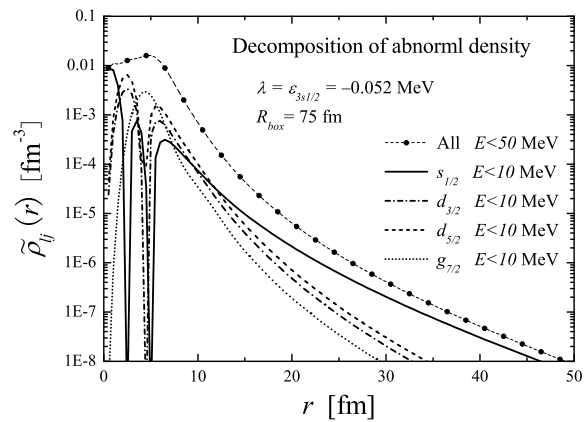}
\vspace{-8mm}
  \caption{
The abnormal density $\tilde{\rho}(r)$ and
the contributions from the low-lying resonant states 
$\tilde{\rho}_{lj}(r)$ 
obtained by the Woods-Saxon potential plus HFB pairing model 
in the case of 
$\lambda$$=$$\varepsilon_{3s_{1/2}}$$=$$-0.052$ MeV.
The box size is fixed to be 75 fm.
}
\label{FIG_arho}
\end{figure}   

\begin{figure}[tb]
\center
\includegraphics[scale=0.83]{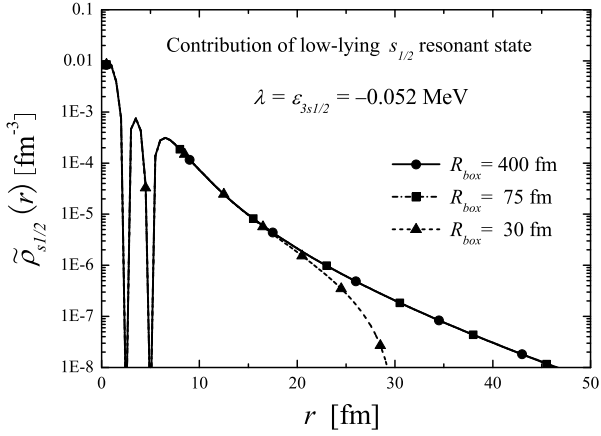}
\vspace{-8mm}
  \caption{The box size dependence of $\tilde{\rho}_{s_{1/2}}(r)$
in the case of 
$\lambda$$=$$\varepsilon_{3s_{1/2}}$$=$$-0.052$ MeV.
The results obtained with $R_{box}=$ 30, 75, and 400 fm are compared.
}
\label{FIG_s-arho}
\end{figure}

In the present subsection, we discuss the pairing correlation
in loosely-bound nuclei.
We focus on the applicability of HFB calculation with 
the box boundary condition for the extreme situation;
the pairing effect on the $s_{1/2}$ state 
in the limit of $\varepsilon_{s_{1/2}} \to 0$. 
In Fig.\ref{FIG_Epair} the average pairing gap
and the pairing energy around $^{86}$Ni are plotted
as a function of $\lambda=\varepsilon_{3s_{1/2}}$.
The single-particle energy changes by varying
$V_{WS}$ while keeping the other parameters fixed.
The box boundary condition with $R_{box}=75$ fm is imposed.
Pairing correlation becomes monotonically stronger
as $\varepsilon_{3s_{1/2}}$ increasing up to around -0.5 MeV,
because of the change of the shell structure
that is self-consistently taken into account
in the pairing potential.
The pairing correlation is much enhanced in the
very loosely-bound region, $\lambda=\varepsilon_{3s_{1/2}}$
$>$ $-0.5$ MeV, due to the increasing coupling to 
the nearby continuum \cite{BD99}.
The importance of the coupling to the non-resonant continuum states 
for pairing correlation
is also discussed in connection with di-neutron correlation
\cite{MM04}.

To examine whether such enhancement of pairing correlation
is due to the artifact of the box boundary condition, 
we examined the box size dependence up to $R_{box}=400$ fm.
In Fig.\ref{FIG_Epair-cnv} the pairing energy in the case of
$\lambda=\varepsilon_{3s_{1/2}}=-0.052$ MeV is shown as a function
of $R_{box}$. 
The strength $V_{WS}$ is fixed to be -39.0 MeV.
We confirm that the box size 75 fm is enough
to describe the pairing correlation in such very loosely-bound nuclei.

To estimate the pairing effect in the $3s_{1/2}$ state,
Hamamoto {\it et al.} \cite{HS04} introduced 
the effective pairing gap by
$\Delta_{eff}(s_{1/2}) \equiv E_{s_{1/2}}$,
where the smallest quasiparticle energy $E_{s_{1/2}}$ is calculated
with the condition $\lambda=\varepsilon_{3s_{1/2}}$
for the discrete state satisfying $(\lambda+E_{s_{1/2}})<0$.
In the case of $(\lambda+E_{s_{1/2}})>0$, the value of
$E_{s_{1/2}}$ is obtained by calculating the derivative of
the phase sift of the quasiparticle wave function
$u_{s_{1/2}}(E_{s_{1/2}},r)$.
In our analysis, to estimate the average pairing effect 
on the resonant state,
we introduce the average effective pairing gap by
\begin{eqnarray}
\Delta_{eff}(s_{1/2})=E_{av}(s_{1/2}),
\end{eqnarray}
where $E_{av}(s_{1/2})$ is the average energy of the resonant
state with the occupation probability,
\begin{eqnarray}
E_{av}(s_{1/2})&=&\frac{1}{N_{res}(s_{1/2})}  \nonumber \\
& & \hspace{-10mm} \times \sum_{n} \theta (E_{s_{1/2},n}<10) 
(v_{s_{1/2},n})^2 E_{s_{1/2},n}, 
\label{EQ-DELeff-AV}
\end{eqnarray}
where $\theta(x<a)=$1 for $x<a$, otherwise 0.
The occupation probability at $E_{lj,n}$
is defined by
\begin{eqnarray}
(v_{lj,n})^2=\int_{0}^{R_{box}} \{v_{lj}
(E_{lj,n},r)\}^2 dr.
\end{eqnarray}
The integrated occupation probability 
over the resonant state, 
\begin{eqnarray}
N^{(res)}_{lj}=
\sum_{n} \theta (E_{lj,n}<10) (v_{lj,n})^2,
\end{eqnarray}
is introduced.

In Fig.\ref{FIG_DELeff} $\Delta_{eff}(s_{1/2})$
is plotted as a function of $\lambda=\varepsilon_{3s_{1/2}}$.
As $\lambda=\varepsilon_{3s_{1/2}}$ increases, 
$\Delta_{eff}(s_{1/2})$ monotonically increases.
The enhancement of $\Delta_{eff}(s_{1/2})$ 
in the limit of $\lambda=\varepsilon_{3s_{1/2}} \to 0$ appears.
This behavior is consistent with
the pairing correlation of the total system measured by
$\bar{\Delta}$ and $E_{pair}$ (see Fig.\ref{FIG_Epair}).
This means that the strong coupling between the pairing potential 
and the $s_{1/2}$ state remains in the limit of
$\lambda=\varepsilon_{3s_{1/2}} \to 0$.

We also examined the box size dependence of $\Delta_{eff}(s_{1/2})$.
In Fig.\ref{FIG_s-ocp} the occupation probability of the $s_{1/2}$
state in the case of $\lambda=\varepsilon_{3s_{1/2}}=-0.052$ MeV
is shown as a function of $E_{s_{1/2}}$.
The results obtained with $R_{box}=$ 75 and 400 fm are compared.
Although the level density of the discretized continuum states
is very different, the average structure of the resonance is unchanged.
Namely the integrated occupation probability $N^{(res)}_{s_{1/2}}$ 
is 0.430 (0.429), and the average effective pairing gap 
$\Delta_{eff}(s_{1/2})$ is
1.15 (1.13) MeV with $R_{box}=75$ (400) fm.
And also, as we will discuss in the next subsection, 
the average root-mean-square (rms) radius of the lower component 
of the $s_{1/2}$ state is independent of $R_{box}$
larger than 50 fm (see Fig.\ref{FIG_Epair-cnv}).

To emphasize the importance of self-consistency in
the pairing potential, 
we perform HFB calculation
with the analytic surface-type pairing potential \cite{HM03},
\begin{eqnarray}
\Delta (r) \varpropto r\frac{df(r)}{dr},
\label{EQ-WSPAIR}
\end{eqnarray}
where $f(r)$ is the Woods-Saxon type function of Eq.(\ref{EQ-WSFORM}).
The average pairing gap with the analytic potential is
defined by 
\begin{eqnarray}
\bar{\Delta}=\frac{\int_0^{R_{box}}\Delta(r)f(r)r^2 dr}
{\int_0^{R_{box}}f(r)r^2 dr}.
\end{eqnarray}

In Fig.\ref{FIG_DELeff} the average effective pairing gap 
$\Delta_{eff}(s_{1/2})$ obtained by HFB with 
the analytic pairing potentials
of $\bar{\Delta}=1.0$ and $1.5$ MeV are also shown.
These  $\Delta_{eff}(s_{1/2})$ decrease rapidly in the limit of
zero binding energy, due to the decoupling between the $s_{1/2}$ state
and the pairing potential as discussed in Ref.\cite{HM04}, 
contrary to the HFB calculation with 
the self-consistent pairing potential.

The low-lying $s_{1/2}$ state
is the dominant component of $\tilde{\rho}(r)$ 
outside the surface region in the HFB calculation with
the self-consistent pairing potential.
To see this, we define the contributions from the low-lying resonant 
states to $\tilde{\rho}(r)$ by
\begin{eqnarray}
\tilde{\rho}_{lj}(r)&=&-\frac{(2j+1)}{4\pi r^2} 
\sum_{n} 
\theta (E_{lj,n}<10)  \nonumber \\
& &\quad \times u_{lj}(E_{lj,n},r) v_{lj}(E_{lj,n},r).
\end{eqnarray}
In Fig.\ref{FIG_arho}, $\tilde{\rho}_{lj}(r)$
are shown in the case of $\lambda=\varepsilon_{3s_{1/2}}=-0.052$ MeV
with $R_{box}=75$ fm.
As clearly seen, the $\tilde{\rho}_{s_{1/2}}(r)$
has a sizable component outside the surface region, $r>10$ fm.
In Fig.\ref{FIG_s-arho} the box size dependence of
$\tilde{\rho}_{s_{1/2}}(r)$ is examined, and we confirmed
that $R_{box}=$ 75 fm is enough to describe the pairing correlations
in such very loosely-bound nuclei.
Because of the spatially extended component in the self-consistent
pairing potential,
the pairing field can act on the $s_{1/2}$ state,
although the loosely-bound $s_{1/2}$ state is spatially extended.
The coupling between the self-consistent pairing potential and 
the loosely-bound $s_{1/2}$ state becomes much stronger
by repeating the process self-consistently 
because of the pairing anti-halo effect 
that we will discuss in the next subsection.
On the other hand, this process is absent in the analytic pairing 
potential, 
and the coupling to the loosely-bound $s_{1/2}$ state becomes
artificially weaker for  $\varepsilon_{3s_{1/2}} \to 0$.

The enhanced pairing energy in the very loosely-bound nuclei
originates from the spatially extended component of $\tilde{\rho}(r)$.
As shown in Fig.\ref{FIG_Epair-cnv} the pairing energy
in the case of $\lambda=\varepsilon_{3s_{1/2}}=-0.052$ MeV
obtained with $R_{box}=400$ (30) fm is -20.368 (-20.238) MeV.
The difference 130 keV is due to the spatially extended
component of $\tilde{\rho}(r)$ shown in Fig.\ref{FIG_s-arho}.


\subsection{Pairing anti-halo effect in lower components}
\label{SUBSEC-PAHE}

\begin{figure}[tb]
\center
\includegraphics[scale=0.75]{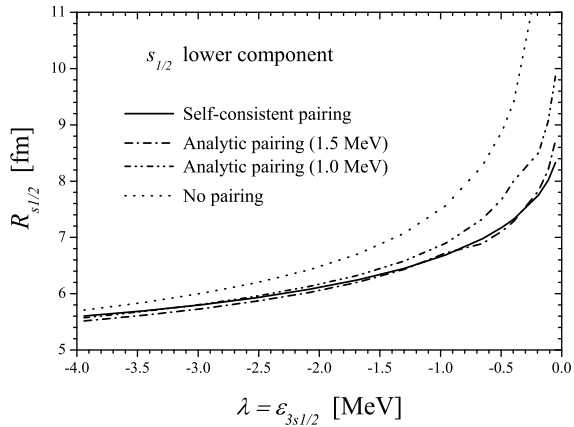}
\vspace{-8mm}
  \caption{The average rms radius of the lower component of
the $s_{1/2}$ quasiparticle wave function 
as a function of $\lambda$$=$$\varepsilon_{3s_{1/2}}$
obtained by the Woods-Saxon potential plus HFB pairing model.
The results with the self-consistent pairing potential
and the analytic pairing potentials of
$\bar{\Delta}=1.0$ and $1.5$ MeV are compared.
The radius without pairing is also shown.
The box size is fixed to be 75 fm.
}
\label{FIG_s-R}
\end{figure}   

\begin{table}
\caption{The quasiparticle energies of the low-lying resonant states
$E^{(res)}_{lj}$ 
obtained by the Woods-Saxon potential plus HFB pairing 
model with $V_{WS}=-41.4$ MeV.
The average rms radii of the lower components $R_{lj}$ and
the integrated occupation probabilities
$N^{(res)}_{lj}$ and $N_{lj}(1.0)$ are shown.
The single-particle energies of the corresponding 
bound or resonant states $\varepsilon_{lj}$ are also listed.
The box size is fixed to be 75 fm.
}
\begin{ruledtabular}
\begin{tabular}{ccccccc}
  & $E^{(res)}_{lj}$ (MeV) & $R_{lj}$ (fm)& $N^{(res)}_{lj}$ & 
$N_{lj}(1.0)$  & $\varepsilon_{lj}$ (MeV)\\ 
\hline
$s_{1/2}$  & 0.80 & 7.18 & 0.48 & 0.37 & -0.50 \\
$d_{3/2}$  & 1.25 & 6.12 & 0.26 & 0.22 & 0.14 \\
$d_{5/2}$  & 1.72 & 5.79 & 0.86 & 0.83 & -1.77 \\  
$g_{7/2}$  & 2.05 & 5.38 & 0.13 & 0.13 & 0.98 
\end{tabular}
\end{ruledtabular}
\label{TAB-QPS}
\end{table}

\begin{figure}[tb]
\center
\includegraphics[scale=0.75]{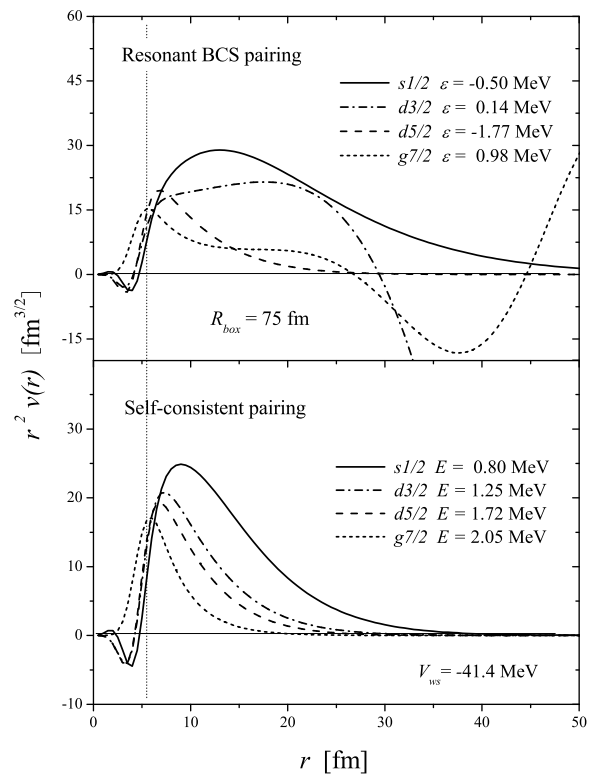}
\vspace{-8mm}
  \caption{The effective hole wave functions
of the $s_{1/2}$, $d_{3/2}$, $d_{5/2}$ and $g_{7/2}$ resonant states 
obtained by the Woods-Saxon potential plus HFB pairing model
are shown.
The functions in the resonant BCS approximation are also plotted.
The strength $V_{WS}=-41.4$ MeV and 
the box size $R_{box}=75$ fm are fixed.
}
\label{FIG_r2v}
\end{figure}

Self-consistent pairing correlation in HFB has
a unique role that changes
the spatial structure of the quasiparticle wave functions
in loosely-bound nuclei.
In the present subsection, we focus on the spatial structure
of the lower component, and the structure of the upper component
will be discussed in the next subsection.

According to the asymptotic HFB equation 
with $V_{lj}(r,r') \to 0$ and $\Delta(r,r') \to 0$
in the limit of $r \to \infty$,
the lower component of the quasiparticle wave function decays 
exponentially for any quasiparticle energy $E_{lj}>0$, 
\begin{eqnarray}
 v_{lj} (E_{lj},r) \to \exp (- \beta_{lj} r),
\label{EQ-ASYHOLEWF}
\end{eqnarray}
where 
$\beta_{lj} = \sqrt { 2m (E_{lj} - \lambda)/\hbar^2}$.
The HFB quasiparticle energy $E_{lj}$ is well approximated by the 
canonical quasiparticle energy, 
$E_{lj}$ $\simeq$ $E_{lj}^{can}$ $\equiv$ 
$\sqrt{(\epsilon_{lj}^{can}-\lambda)^2 +(\Delta_{lj}^{can})^2}$, 
where $\epsilon_{lj}^{can}$ and $\Delta_{lj}^{can}$ are the canonical 
single-particle energy and the canonical pairing gap
of the corresponding state  \cite{DN96}.
This means that $\beta_{lj}$ stays at finite value
\begin{eqnarray}
\beta_{lj} \to \sqrt { 2m E_{lj}/\hbar^2}
\geq \sqrt { 2m \Delta_{lj}^{can}/\hbar^2} >0
\end{eqnarray} 
with non-zero $\Delta_{lj}^{can}$ for $\lambda \rightarrow 0$.
Therefore 
self-consistent pairing correlation changes the asymptotic behavior of 
$v_{lj} (E_{lj},r)$, and especially acts against
the development of the infinite rms radius that characterizes 
the wave functions of $s$ and $p$ states 
in the limit of vanishing binding energy.
This change of the spatial structure in the lower component
of the quasiparticle wave function is called
"the pairing anti-halo effect" in Ref.\cite{BD00}.

In Fig.\ref{FIG_s-R} the average rms radii of 
the lower component of the $s_{1/2}$ state 
with the self-consistent pairing potential and
the analytic pairing potentials of
$\bar{\Delta}$$=$$1.0$ and $1.5$ MeV
are shown as a function of $\lambda=\varepsilon_{3s_{1/2}}$.
The average rms radius $R_{lj}$ is defined by 
\begin{eqnarray}
& &\left(R_{lj}\right)^2=
\frac{1}{N^{(res)}_{lj}} 
\sum_{n} \theta(E_{lj,n}<10) \nonumber \\
& & \qquad \quad  \times  \int_{0}^{R_{box}} r^2
\{v_{lj}(E_{lj,n},r)\}^2 dr.
\label{EQ-RMSRAV}
\end{eqnarray}
The radius of the $3s_{1/2}$ state without pairing goes to 
infinity for $\varepsilon_{3s_{1/2}} \to 0$.
On the other hand, the divergence is suppressed in HFB.
For $\lambda=\varepsilon_{3s_{1/2}}<-1.5$ MeV,
irrespective of the pairing potential,
the radii obtained by the HFB calculations have a similar behavior.
As approaching the zero binding energy, 
the spatial structure of the lower component becomes
sensitive to the treatment of pairing correlation
in accordance with the behavior of the average effective 
pairing gap $\Delta_{eff}(s_{1/2})$ shown in Fig.\ref{FIG_DELeff}.

In connection with the spatial structure of two-quasiparticle states,
we examine the effective hole wave function of the transition 
with multipolarity $L$ defined by
\begin{eqnarray}
f_{lj,n}^{(L)}(r)=r^L v_{lj}(E_{lj,n},r)/v_{lj,n}.
\end{eqnarray}
The factor $v_{lj,n}(>0)$ is introduced for the normalization.
The effective hole wave functions 
of the $s_{1/2}$, $d_{3/2}$, $d_{5/2}$ and $g_{7/2}$
resonant states with $L=2$ are plotted in Fig.\ref{FIG_r2v}.
The strength $V_{WS}$ is fixed to be -41.4 MeV to simulate 
the neutron shell structure in $^{86}$Ni.
The properties of these resonant states are summarized
in Table.\ref{TAB-QPS}.
In HFB the effective hole wave function is localized
according to Eq.(\ref{EQ-ASYHOLEWF}), irrespective of the
corresponding single-particle energy. 
However the localization depends on $\ell$.
The function $f_{s_{1/2}}^{(2)}(r)$ has the large component 
not only around the surface region, but also the extended region,
10 fm $<r<$ 20 fm.
The component in the extended region remains, but becomes
smaller in the $d_{3/2}$ and $d_{5/2}$ states,
and negligible in the $g_{7/2}$ state.
In Subsec.\ref{SEC-TQPS} we will discuss the impact 
of such spatially extended part of the lower component 
in connection with the new aspects of 
low-frequency vibrational excitations in neutron drip line nuclei.

For comparison, we analyze the effective hole wave functions
in the resonant BCS approximation.
Although the BCS approximation is not well defined for unstable nuclei
due to the unphysical neutron gas \cite{DF84},
recently the extended version, the resonant BCS approximation,
was proposed \cite{SL97,SG00,SG03,CM04}. In this approach, 
resonant states are taken into account for the particle states.
Applicability of the resonant BCS method for ground state properties 
is discussed in Ref.\cite{GS01}.
In this approximation, 
the upper and lower components of the quasiparticle wave function
are simply proportional to the corresponding single-particle 
wave function $\varphi _{lj} (\varepsilon_{lj,n}, r)$,
\begin{eqnarray}
\begin{array}{l}
 u_{lj} (E_{lj,n},r) = u_{lj,n}^{BCS} \,
 \varphi _{lj} (\varepsilon_{lj,n},r), \\ 
 v_{lj} (E_{lj,n},r) = v_{lj,n}^{BCS} \,
 \varphi _{lj} (\varepsilon_{lj,n},r). \\ 
 \end{array}
\label{EQ-BCS-ASY} 
\end{eqnarray}
The quasiparticle energy $E_{lj,n}$ is related to the
single-particle energy $\varepsilon_{lj,n}$ by
\begin{eqnarray} 
E_{lj,n}=\sqrt{(\varepsilon_{lj,n}-\lambda)^2 +(\Delta_{lj,n})^2},
\end{eqnarray}
where the Fermi energy $\lambda$ and the pairing gap $\Delta_{lj,n}$.
$v_{lj,n}^{BCS}$ ($u_{lj,n}^{BCS}$) is
the occupation (unoccupation) amplitude.
The index $n$ runs over not only the bound states but also
the resonant states.
For $r \to \infty$, the radial single-particle wave function of 
the bound state decays exponentially,
\begin{eqnarray}
\varphi_{lj}(\varepsilon _{lj,n},r) \to \exp(-\alpha_{lj,n} r),
\end{eqnarray}
where $\alpha_{lj,n}=\sqrt{-2m \varepsilon_{lj,n} /\hbar^2}$.
On the other hand, the wave function in the continuum region
satisfies the scattering boundary condition,
\begin{subequations}
\begin{eqnarray}
\varphi_{lj} (\varepsilon_{lj} ,r) &\to&
\left[ \cos (\delta _{lj}) r j_l (k_{lj} r)
- \sin (\delta _{lj}) r n_l (k_{lj} r) \right] \label{EQ-OGB} \\
&\sim& \sin (k_{lj} r+\delta'_{lj}) \label{EQ-SBC}
\end{eqnarray}
\end{subequations}
where $k_{lj} = \sqrt{ 2m \varepsilon_{lj}/\hbar^2}$.
$j_l$ and $n_l$ are spherical Bessel and Neumann functions, 
and $\delta_{lj}$ ($\delta'_{lj}\equiv \delta_{lj}-l\pi/2$) is 
the phase shift corresponding to the angular momentum $(lj)$.

As shown in Fig.\ref{FIG_r2v},
the effective hole wave functions of the $d_{3/2}$ and $g_{7/2}$
resonant states in the resonant BCS approximation diverge 
for $r \to \infty$ in contrast with those in HFB.
The effective hole wave functions of the bound single-particle states
in the (resonant) BCS approximation are
essentially the same with those without pairing.
The function for the halo state, $f_{s_{1/2}}^{(2)}(r)$,
is spatially very extended due to the absence of 
the pairing anti-halo effect. On the other hand, 
$f_{d_{5/2}}^{(2)}(r)$ has almost the same
spatial structure with that in HFB, 
because the corresponding single-particle state 
is the bound state without halo.

\subsection{Broadening effect in upper components}

\begin{figure}[tb]
\center
\includegraphics[scale=0.8]{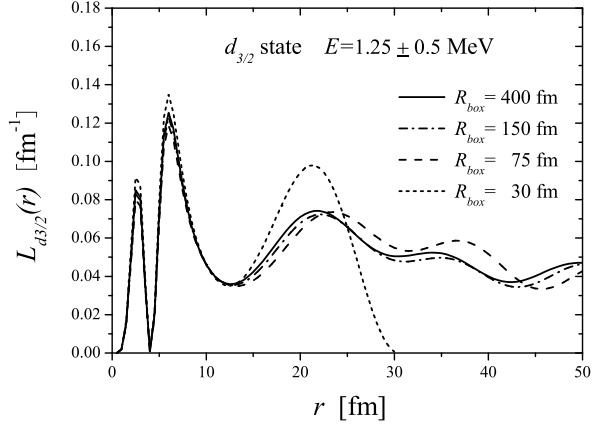}
\vspace{-8mm}
  \caption{The localization indicator of the resonant
$d_{3/2}$ state at $E^{(res)}=1.25$ MeV in the Woods-Saxon potential
plus HFB pairing model is shown.
The results obtained with $R_{box}=$ 30, 75, 150, and 400 fm 
are compared.
}
\label{FIG_L-d3}
\end{figure}   

\begin{figure}[tb]
\center
\includegraphics[scale=0.72]{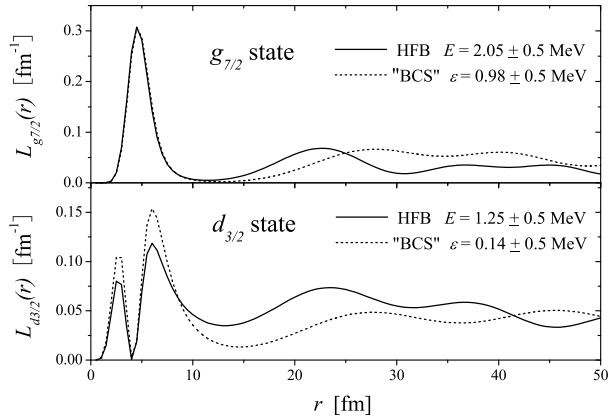}
\vspace{-9mm}
  \caption{The localization indicators of the resonant 
$d_{3/2}$ state at $E^{(res)}=1.25$ MeV 
and the resonant $g_{7/2}$ state at $E^{(res)}=2.05$ MeV 
in the Woods-Saxon potential
plus HFB pairing model are shown.
The "BCS" localization indicators of the corresponding states
are also plotted.
The box size is fixed to be 75 fm.
}
\label{FIG_L-d3-g7}
\end{figure}   

\begin{figure}[tb]
\center
\includegraphics[scale=0.72]{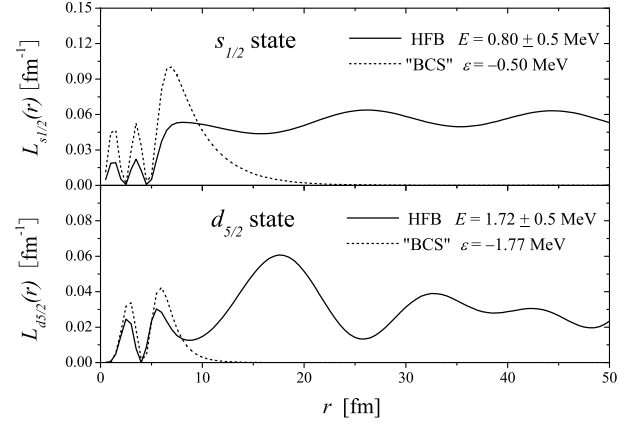}
\vspace{-9mm}
  \caption{The same as Fig.\ref{FIG_L-d3-g7} but 
for the resonant 
$s_{1/2}$ state at $E^{(res)}$=0.80 MeV 
and the resonant $d_{5/2}$ state at $E^{(res)}$=1.72 MeV.
}
\label{FIG_L-s-d5}
\end{figure}

Self-consistent pairing correlation also changes 
the spatial structure of the upper component of
the quasiparticle wave function.
Asymptotic behavior of the upper component
depends on the quasiparticle energy \cite{DF84}.
In the discrete region $(E_{lj,n}+\lambda)<0$, 
the wave function decays exponentially,  
\begin{eqnarray}
 u_{lj} (E_{lj,n}, r) \to \exp (- \alpha_{lj,n} r),
\end{eqnarray}
where
$\alpha_{lj,n} = \sqrt { -2m (E_{lj,n}+\lambda)/\hbar^2}$.
In the continuum region $(E_{lj}+\lambda)>0$, 
on the other hand, 
\begin{subequations}
\begin{eqnarray}
 u_{lj} (E_{lj},r) &\to&
\left[ \cos \left( \delta_{lj}  \right) 
r j_l ( k_{lj} r ) - 
\sin \left( \delta_{lj} \right) 
r n_l ( k_{lj} r) \right], \\
&\sim& \sin (k_{lj} r +\delta'_{lj}),
\label{EQ-ABPWF}
\end{eqnarray}
\end{subequations}
with $k_{lj}  = \sqrt { 2m (E_{lj} + \lambda) /\hbar^2}$.

In the neutron drip line region with $\lambda \to 0$,
all quasiparticle states are in the continuum region.
To examine the spatial structure of the upper components,
we introduce the localization indicator,
\begin{eqnarray}
L_{lj}(r;\delta E)=\sum_{n} \theta (|E_{lj,n}-E_{lj}^{(res)}|
<\delta E/2) 
\{u_{lj}(E_{lj,n},r)\}^2.
\end{eqnarray}
The summation is taken around the resonance energy
$E_{lj}^{(res)}$ within the energy window $\delta E$.
In principle, the localization indicator is independent of
$R_{box}$ with arbitrary $\delta E$, if
$R_{box}$ is enough large.
In Fig.\ref{FIG_L-d3} the indicator of the $d_{3/2}$ state
at $E^{(res)}=1.25$ MeV obtained
with $R_{box}$ $=$ 30, 75, 150, and 400 fm are compared.
The energy window $\delta E$ is fixed to be 1.0 MeV.
Apart from the artifact around the box boundary, 
these results coincide in good accuracy.
The box size $R_{box}=75$ fm is enough for
our discussion on the spatial structure of the two-quasiparticle
states around $^{86}$Ni, because it is required 
to describe the spatial structure of the upper components within 30 fm
as seen in Fig.\ref{FIG_r2v}.

In Fig.\ref{FIG_L-d3-g7} the indicators of the $d_{3/2}$ state
at $E^{(res)}=$1.25 MeV and the $g_{7/2}$ state at
$E^{(res)}=$2.05 MeV, that originate from the resonant 
single-particle states, are shown.
The occupation probabilities within the energy window,
\begin{eqnarray}
N_{lj}(\delta E)=\sum_{n} \theta(|E_{lj,n}-E_{lj}^{(res)}|<\delta E/2) 
(v_{lj,n})^2,
\end{eqnarray}
together with $\delta E=1$ MeV are listed in Table.\ref{TAB-QPS}.
To examine the role of self-consistent pairing correlation,
as the reference distribution,
the localization indicator in the resonant BCS approximation 
(we call the "BCS" localization indicator)
for $\varepsilon_{lj}^{(res)}>0$ is introduced  by
\begin{eqnarray}
& &L_{lj}(r;\delta E)=(1-N_{lj}(\delta E)) \times \nonumber \\
& & \quad \sum_{n} 
\theta (|\varepsilon_{lj,n}-\varepsilon_{lj}^{(res)}|<\delta E/2) 
\{\varphi_{lj}(\varepsilon _{lj,n},r)\}^2.
\label{EQ-LBCSP}
\end{eqnarray}
The single-particle wave function 
$\varphi_{lj}(\varepsilon _{lj,n},r)$ with $\varepsilon _{lj,n}>0$
is the solution of the Woods-Saxon potential of Eq.(\ref{EQ-WSPOT})
with  the box boundary condition.

In the $g_{7/2}$ resonant state, the difference
between the HFB and "BCS" indicators is small within 20 fm.
On the other hand, in the $d_{3/2}$ state,
the spatial localization of the resonance is weakened in HFB;
namely, the component within the centrifugal barrier
($r<10$ fm) is reduced, and enhanced in the extended region
(10 fm $<r<$ 20 fm).
This is the direct manifestation of the strong mixing with
the nearby non-resonant continuum states.
As discussed in Ref.\cite{DF84,HM03,GS01,FT00},
the resonance width of the quasiparticle state originating
from the resonant (bound) single-particle state 
is broadened (acquired) in HFB. 
This effect is more prominent in lower-$\ell$ states
with smaller quasiparticle energy.
Consequently, the spatial structure of
the upper component of the quasiparticle wave function changes, 
and the spatial localization is weakened
by the mixing with continuum states.
To emphasize this effect induced 
by self-consistent pairing correlation;
the broadened resonance width and the spatially weakened localization,
we call "the broadening effect" in the upper component.

In Fig.\ref{FIG_L-s-d5} the localization indicators of 
the $s_{1/2}$ state at $E^{(res)}=0.80$ MeV
and the $d_{5/2}$ state at $E^{(res)}=1.72$ MeV,
that originate from the bound single-particle states, are shown.
The "BCS" localization indicator
with $\varepsilon _{lj,n}<0$ is defined by
\begin{eqnarray}
& &L_{lj}(r)=(1-N_{lj}(\delta E)) \{\varphi_{lj}(\varepsilon _{lj,n},r)\}^2.
\end{eqnarray}
In these states, the spatial structure of 
the HFB and "BCS" indicators is very different.
The "BCS" indicator is localized, but non-localized in HFB,
because of the different boundary conditions, and
the broadening effect is absent in the (resonant) BCS approximation.
The broadening effect is strong in the upper component
of the $s_{1/2}$ and $d_{3/2}$ states.
Especially the localization around the surface region is
strongly affected in the $s_{1/2}$ state.


\section{Spatial structure of two-quasiparticle states} 
\label{SEC-TQPS}

\subsection{Two-quasiparticle states in HFB}

The spatial localization of two-quasiparticle states
(\opoh states in closed shell nuclei)
is a necessary condition to realize the coherent motions among them.
In stable nuclei the spatial localization of
the contributing two-quasiparticle states is realized 
around the surface region (for example, see Ref.\cite{GM65,FP67}).
In neutron drip line nuclei, on the other hand, 
because the contributing single-particle states are tightly-bound states, 
loosely-bound states, resonant and non-resonant continuum states,
the spatial distributions of the two-quasiparticle states
have rich variety of the spatial character.
Therefore, first of all, the realization of coherent motions
among them is not obvious. 
If they can be realized, we may expect
qualitatively new aspects of low-frequency vibrational
excitations in neutron drip line nuclei.

In the present study, we analyze the transitions
from the ground state to excited states within the same nucleus.
The spatial distribution of the particle-hole component
of the two-quasiparticle state between
the upper component $u_{l'j'} (E_{l'j'},r)$ and
the lower component $v_{lj} (E_{lj},r)$ is defined by
\begin{eqnarray}
F_{uv}^{(L)}(r)\equiv u_{l'j'} (E_{l'j'},r) r^L 
v_{lj} (E_{lj},r)/N_{uv},
\end{eqnarray}
where $N_{uv}$ $=$ $\sqrt{1-(v_{l'j'})^2} v_{lj}$ $(>0)$ 
is the normalization factor.

One of the novel features of low-frequency vibrational excitations
in neutron drip line nuclei is 
that two-quasiparticle states composed of quasiparticles both
in the continuum region are inevitably involved.
In HFB, owing to the pairing anti-halo effect,
the asymptotic behavior for $r \to \infty$ is,
\begin{eqnarray}
F_{uv}^{(L)}(r) \to r^L \sin(k_{l'j'} r +\delta'_{l'j'})
\exp( -\beta_{lj} r).
\label{EQ-TQPC}
\end{eqnarray}
Since $\beta_{lj}$ stay at finite with non-zero pairing correlation, 
the spatial distribution is localized
irrespective of the quasiparticle energy.

Recently QRPA calculation with the resonant BCS approximation
is applied to describe vibrational excitations in loosely-bound nuclei
\cite{HG04,CM05}. 
In this method, according to Eq.(\ref{EQ-SBC}),
the spatial distribution of the two-quasiparticle state
composed of quasiparticles both in the continuum
is non-localized function with the asymptotic behavior,
\begin{eqnarray}
F_{uv}^{(L)}(r) &\to& r^L  \sin (k_{l'j'} r +\delta'_{l'j'}) 
\sin (k_{lj} r +\delta'_{lj}).
\label{EQ-BCSPP}
\end{eqnarray}
Consequently the spatial integration, 
namely the transition matrix element, diverges. 
In practice this divergence may be suppressed by imposing the box 
boundary condition with small box size, 
typically $R_{box}=15$ fm \cite{HG04}.
However a large box radius is needed to represent 
the wave functions with the spatially extended structure
in good accuracy.
Therefore this approximation is uncontrollable 
for the description of excitations in nuclei 
close to the neutron drip line.
In Ref.\cite{CM05} the low-lying dipole modes in $^{26,28}$Ne 
are studied by means of quasiparticle relativistic RPA 
with the resonant BCS approximation.
In this calculation, although the resonant states are calculated by 
imposing the scattering boundary condition, the obtained results
don't suffer with the problem of the divergence because of
the peculiarity of the shell structure.
For example, the two-quasiparticle state between 
the resonant $f_{7/2}$ and $p_{3/2}$ states   
doesn't contribute to the dipole excitation in $^{28}$Ne.
Moreover the unoccupied states outside of the pairing
active space are obtained by expanding on a set of the harmonic
oscillator basis, and the divergence of the two-quasiparticle states
between the occupied part of the resonant states and the non-resonant
continuum states is artificially eliminated.


\subsection{Broad localization of two-quasiparticle states}
\label{SUBSEC-TQP}

\begin{figure}[tb]
\center
\includegraphics[scale=0.75]{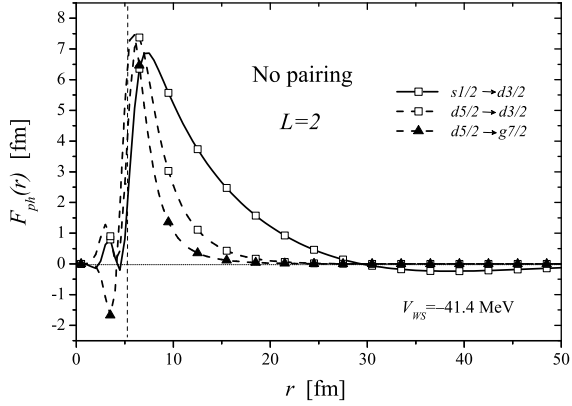}
\vspace{-8mm}
  \caption{The $F_{ph}^{(2)}(r)$ between
the bound states and the resonant states with
$\varepsilon_p$$-$$\varepsilon_h$$<$$5$ MeV in $^{86}$Ni.
}
\label{FIG_Fph}
\end{figure}

\begin{figure}[tb]
\center
\includegraphics[scale=0.68]{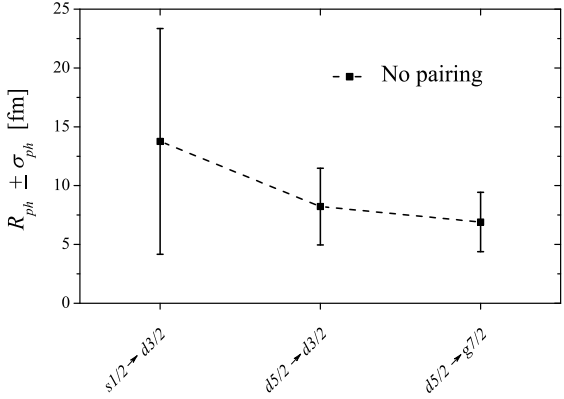}
\vspace{-8mm}
  \caption{The first moments $R_{ph}^{(2)}$
of $F_{ph}^{(2)}(r)$ in Fig.\ref{FIG_Fph}.
The regions $|R_{ph}^{(2)} - r| < \sigma_{ph}^{(2)}$ 
are shown by vertical line segments.
}
\label{FIG_Rph}
\end{figure}   

\begin{figure}[tb]
\center
\includegraphics[scale=0.75]{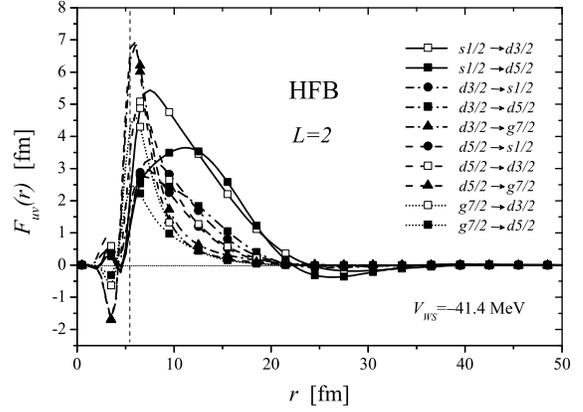}
\vspace{-8mm}
  \caption{The $F_{uv}^{(2)}(r)$ associated with
the low-lying $s_{1/2}$, $d_{3/2}$, $d_{5/2}$, and $g_{7/2}$ 
resonant states in $^{86}$Ni obtained by 
the Woods-Saxon potential plus HFB pairing model. 
}
\label{FIG_Fuv}
\end{figure}   

\begin{figure}[tb]
\center
\includegraphics[scale=0.68]{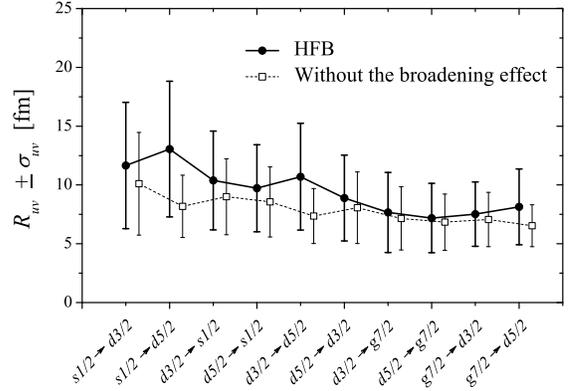}
\vspace{-8mm}
  \caption{
The first moments $R_{uv}^{(2)}$ 
of $F_{uv}^{(2)}(r)$ in Fig.\ref{FIG_Fuv}.
The regions $|R_{uv}^{(2)} - r| < \sigma_{uv}^{(2)}$ 
are shown by vertical line segments.
The result without the broadening effect in the upper components
of the quasiparticle wave functions is also shown.
}
\label{FIG_Ruv}
\end{figure}   

As the illustrative example of neutron drip line nuclei, 
we examine the spatial structure of the neutron quadrupole
two-quasiparticle states in $^{86}$Ni.
To characterize the spatial structure of $F_{uv}^{(L)}(r)$,
we introduce the first moment,
\begin{eqnarray}
R_{uv}^{(L)}=\frac{\int_{0}^{R_{box}} r |F_{uv}^{(L)}(r)|dr}
{\int_{0}^{R_{box}} |F_{uv}^{(L)}(r)|dr},
\end{eqnarray}
and the variance,
\begin{eqnarray}
\left(\sigma_{uv}^{(L)}\right)^2=\frac{\int_{0}^{R_{box}} r^2 
|F_{uv}^{(L)}(r)|dr}
{\int_{0}^{R_{box}} |F_{uv}^{(L)}(r)|dr} - 
\left(R_{uv}^{(L)}\right)^2.
\end{eqnarray}
If the regions $|R_{uv}^{(L)} - r| < \sigma_{uv}^{(L)}$ 
have overlap, we expect the correlations 
among the two-quasiparticle states.

In Fig.\ref{FIG_Fph} the spatial distributions 
without pairing, $F_{ph}^{(2)}(r)$, are shown. 
Because the available particle states are only
the $d_{3/2}$ and $g_{7/2}$ resonant states, 
the \opoh excitations to the resonant states
with $\varepsilon_p -\varepsilon_h < 5$ MeV are
(a) $3s_{1/2}$ $\to$ resonant $d_{3/2}$,
(b) $2d_{5/2}$ $\to$ resonant $d_{3/2}$, and
(c) $2d_{5/2}$ $\to$ resonant $g_{7/2}$.  
The corresponding $R_{ph}^{(2)}$ and the regions
$|R_{ph}^{(2)} - r| < \sigma_{ph}^{(2)}$ are shown 
in Fig.\ref{FIG_Rph}.
These configurations have
overlap around the surface region, and the correlations among them
bring some low-lying collectivity.
However the lowest \opoh state (a) only have the appreciable
component outside the surface region that can't correlate with
the other configurations, and
the lowest RPA solution becomes a single-particle like
excitation as we will discuss
in Subsec.\ref{SUBSEC-TWOPLUS}.

In Fig.\ref{FIG_Fuv}
the spatial distribution of the two-quasiparticle states
associated with the low-lying 
$s_{1/2}$, $d_{3/2}$, $d_{5/2}$ and $g_{7/2}$ resonant states
are plotted.
Due to the pairing anti-halo effect, all distribution functions
are localized in HFB. 
The corresponding $R_{uv}^{(2)}$ and 
the regions $|R_{uv}^{(2)} - r| < \sigma_{uv}^{(2)}$
are shown in Fig.\ref{FIG_Ruv}.
The distributions of the two-quasiparticle states between
the $s_{1/2}$, $d_{3/2}$ and $d_{5/2}$ resonant states 
have the sizable components 
in the spatially extended region around 10 fm $<r<$ 15 fm, 
in addition to the surface region 
where the localization is achieved in stable nuclei.
Because of the spatially extended structure, 
the transition matrix elements
can be larger than those of the tightly bound states in stable nuclei.
Therefore we can expect the large collectivity of low-frequency
vibrational excitations by the spatially extended coherence 
among the two-quasiparticle states.
We call such spatially extended distribution of
the two-quasiparticle states "the broad localization".

In Figs.\ref{FIG_Fuv} and \ref{FIG_Ruv} the spatial structure
of the two-quasiparticle states involving the $g_{7/2}$ resonant state
is also shown.
Because the upper and lower components of the $g_{7/2}$ 
quasiparticle state are strongly confined within the centrifugal 
barrier, the $F_{uv}^{(2)}(r)$ concentrate 
only around the surface region.
In general, we can expect the large collectivity
by the broad localization with low-$\ell$ neutrons, 
and the effect is weaker in the two-quasiparticle states
involving larger-$\ell$ states.

The broad localization is realized 
owing to the broadening effect in the upper component 
of the quasiparticle wave function. 
For comparison, the $R_{uv}^{(2)}$ and 
the regions $|R_{uv}^{(2)} - r| < \sigma_{uv}^{(2)}$
 without the broadening effect, that are calculated with
the lower components obtained by HFB and the upper components 
obtained by the resonant BCS approximation, 
are shown in Fig.\ref{FIG_Ruv}. 
If the broadening effect is absent, the distributions 
concentrate only around the surface region
with small $\sigma_{uv}^{(2)}$.
In this situation, the collective motions among them can appear,
however the low-frequency vibrational excitations have a similar 
behavior with those in stable nuclei, and novel features
of neutron drip line nuclei are suppressed.


\section{HFB plus QRPA calculation} \label{SEC-QRPA}

\subsection{Formulation and inputs}

\begin{figure}[tb]
\center
\includegraphics[scale=0.75]{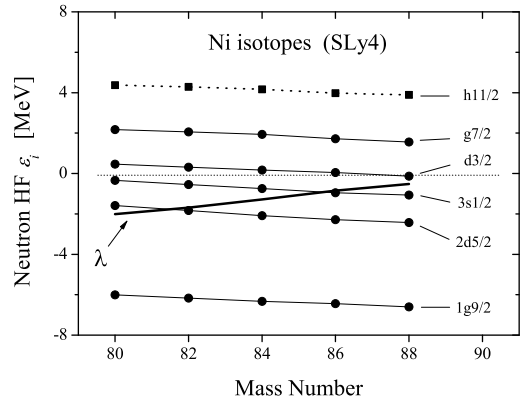}
\vspace{-5mm}
  \caption{The neutron single-particle energies in
neutron rich Ni isotopes obtained by HF with Skyrme SLy4 force. 
The approximate resonance energies with $R_{box}=30$ fm are also shown.
The neutron Fermi energies in HFB are indicated by thick line.
}
\label{FIG_SPE}
\end{figure}

We consider the first $2^{+}$ states in 
neutron rich Ni isotopes 
to examine the unique role of self-consistent pairing
correlation in the neutron drip line nuclei.
In Fig.\ref{FIG_SPE} the neutron single-particle
energies obtained by HF calculation with Skyrme SLy4 force 
\cite{CB98} are shown.
The neutron drip line nucleus is $^{86}$Ni within the HF calculation.
By taking into account pairing correlations, more neutrons can bound.
However the predicted position of the neutron drip line 
depends on the treatment of pairing correlations,
for example, the drip line nucleus is $^{88}$Ni in Ref.\cite{GS01}
and $^{92}$Ni in Ref.\cite{MD00}. 
The predicted drip line also depends 
on the effective interactions and the frameworks;
namely relativistic or 
non-relativistic approaches (for example, see Ref.\cite{MD00,Me98}). 
In the present study we consider up to $^{88}$Ni,
because our purpose is to investigate the qualitative aspects of 
vibrational excitations in neutron drip line nuclei.

We perform HFB plus QRPA calculation with Skyrme force.
The HFB mean fields are 
determined self-consistently from an effective force 
and the residual interaction of the QRPA problem is derived from the same
force. The QRPA problem is solved by the response function method 
in coordinate space. A detailed account of the method can be found in
Ref.\cite{YG04,YK03,KS02}. 
The residual interaction has an explicit momentum dependence, and
these momentum dependence are explicitly treated in our QRPA calculations.
Because we calculate only natural parity (non spin-flip) excitations, 
we drop the spin-spin part of the residual interaction. 
The residual Coulomb and residual spin-orbit interactions are also dropped.

The ground states are obtained by Skyrme-HFB calculation.
The HFB equation is diagonalized on a Skyrme-HF basis represented
in coordinate space with the box boundary condition 
of $R_{box}=30$ fm. 
Spherical symmetry is imposed on the quasiparticle wave functions.
The quasiparticle states with the energy below 50 MeV
and the orbital angular momentum up to $7\hbar$
are taken into account.
The Skyrme SLy4 force is used for the HF mean-field.
The density-dependent zero-range pairing interaction 
of Eq.(\ref{EQ-DDPI}) with the density $\rho(r)$ of protons
plus neutrons is adopted for the pairing field.  
The parameters are taken to be $V_{pair}=-555$ MeV fm$^{-3}$ and
$\rho_c=0.16 (\approx \rho(0))$ fm$^{-3}$, 
that give the average neutron pairing gap 
$\bar{\Delta}_n \approx 12/\sqrt{86}$ in $^{86}$Ni.

\subsection{First $2^+$ states in Ni isotopes}
\label{SUBSEC-TWOPLUS}

\begin{figure}[tb]
\center
\includegraphics[scale=0.7]{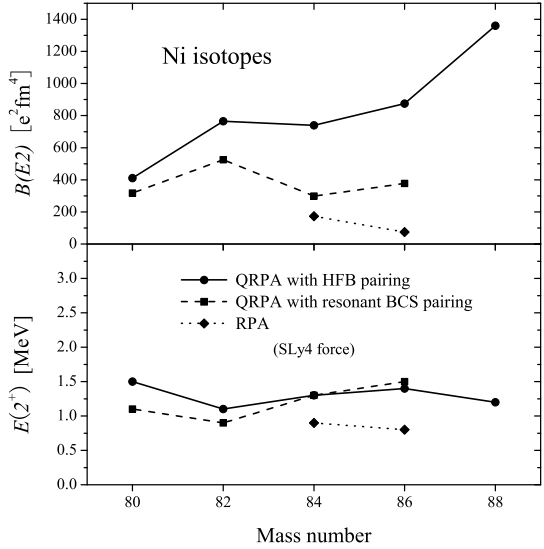}
\vspace{-5mm}
  \caption{The \betwo values and the excitation energies
of the first \twop states in neutron rich Ni isotopes 
obtained by HFB plus QRPA, 
resonant BCS plus QRPA, and RPA
with Skyrme SLy4 force.}
\label{FIG_Ni}
\end{figure}

\begin{figure}[tb]
\center
\includegraphics[scale=0.75]{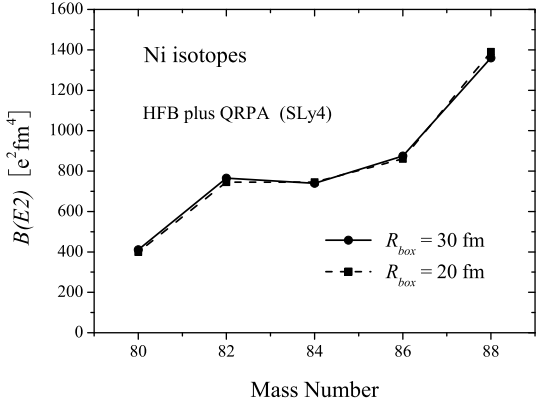}
\vspace{-5mm}
  \caption{The box size dependence of the $B(E2)$ values 
obtained by HFB plus QRPA with Skyrme SLy4 force.}
\label{FIG_CNV-QRPA}
\end{figure}   

\begin{figure}[tb]
\center
\includegraphics[scale=0.75]{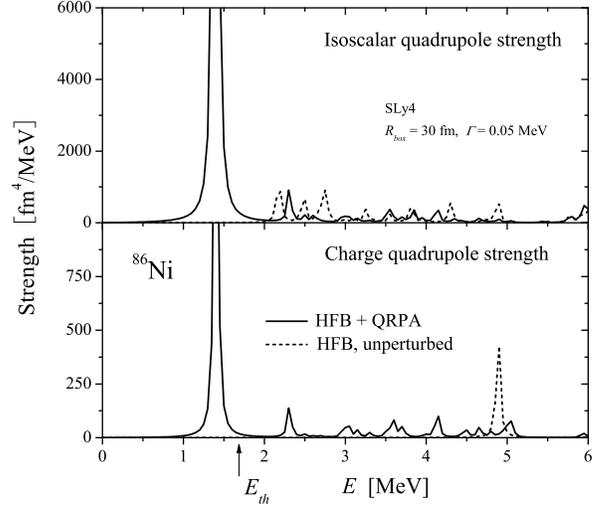}
\vspace{-10mm}
  \caption{The isoscalar and charge quadrupole strength functions 
in  $^{86}$Ni obtained by HFB plus QRPA with Skyrme SLy4 force. 
The corresponding unperturbed strengths are also shown.
A width parameter $\Gamma=0.05$ MeV is introduced for the plots.
The threshold energy $E_{th}$ is indicated by arrow.
}
\label{FIG_QRPA}
\end{figure}   

\begin{figure}[tb]
\center
\includegraphics[scale=0.75]{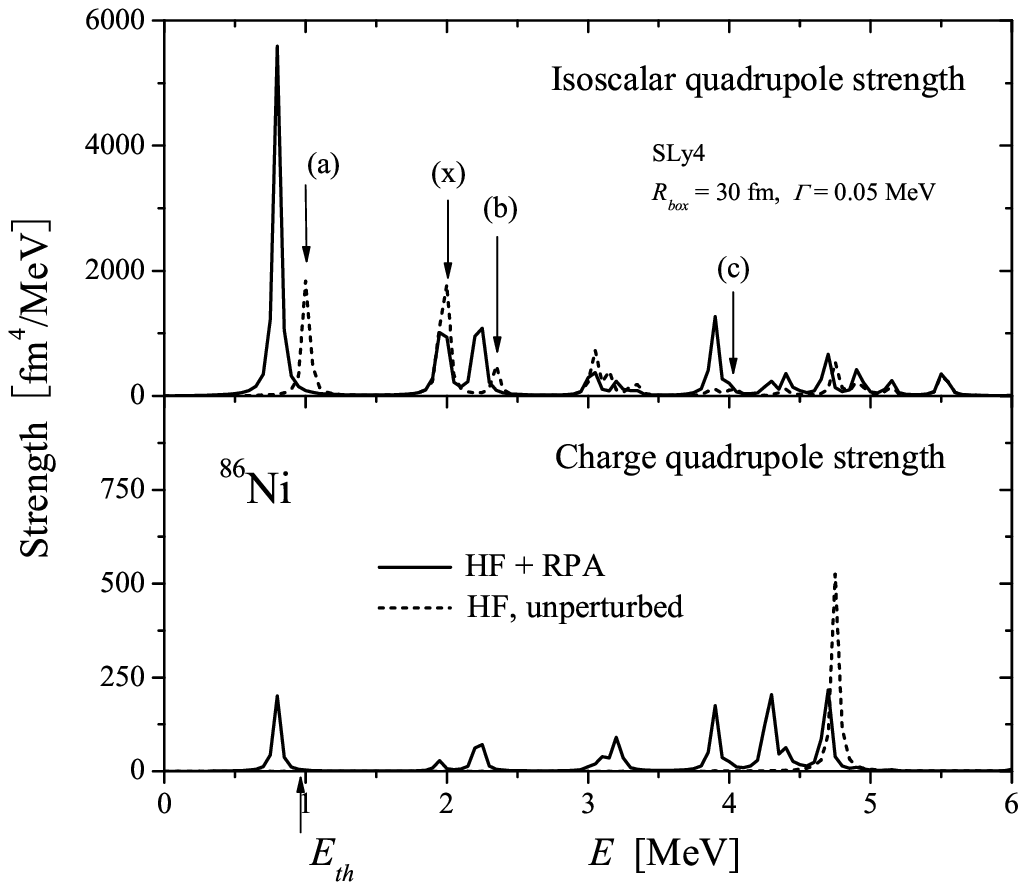}
\vspace{-10mm}
  \caption{The isoscalar and charge quadrupole strength functions 
in  $^{86}$Ni obtained by RPA with Skyrme SLy4 force. 
The corresponding unperturbed strengths are also shown.
The neutron \opoh excitations;
(a) $3s_{1/2}$ $\to$ resonant $d_{3/2}$,
(b) $2d_{5/2}$ $\to$ resonant $d_{3/2}$, 
and
(c) $2d_{5/2}$ $\to$ resonant $g_{7/2}$, 
are indicated by vertical lines.
The configuration,
(x) $3s_{1/2}$ $\to$ (discretized) non-resonant continuum $d_{5/2}$, 
is also pointed out.
A width parameter $\Gamma=0.05$ MeV is introduced for the plots.
The threshold energy $E_{th}$ is indicated by arrow.
}
\label{FIG_RPA}
\end{figure}

In Fig.\ref{FIG_Ni} the \betwo values
and the excitation energies of the first \twop states 
in neutron rich Ni isotopes obtained 
by HFB plus QRPA calculation are shown.
These $2^+$ states are discrete solutions below the threshold energies.
The $B(E2)$ values increase as approaching the neutron drip line.
The $B(E2)$ values in the single-particle unit (=5 times 
the Weisskopf unit) are 7.8 in $^{86}$Ni, and 11.7 in $^{88}$Ni.
To examine the convergence of the solutions, 
the box size dependence of the $B(E2)$ values is shown
in Fig.\ref{FIG_CNV-QRPA}.
Because the spatial distribution
of the contributing two-quasiparticle states is localized
by the pairing anti-halo effect,
the coherence among them occurs within 20 fm 
as shown in Fig.\ref{FIG_Ruv}.
Therefore $R_{box}=20$ fm is enough for the convergence.

In Fig.\ref{FIG_QRPA} the isoscalar and charge quadrupole 
response functions in $^{86}$Ni obtained by HFB plus QRPA, 
and the corresponding unperturbed strength functions are shown.
For the plots, a small width parameter of $\Gamma=0.05$ MeV
is introduced.
All two-quasiparticle states are in the continuum region,
and the configurations below 6 MeV are of neutrons
except for the proton $(2p_{3/2})(1f_{7/2})^{-1}$ configuration
at 4.9 MeV. The strong coherence among the neutrons,
and the successive isoscalar coherent motion of
protons and neutrons lead to the large $B(E2)$ value.

To examine the role of self-consistent pairing correlation, 
the results of the resonant BCS plus QRPA calculation up to $^{86}$Ni
are also plotted in Fig.\ref{FIG_Ni}.
Although the resonant BCS approximation is not well-defined
in neutron drip line region,
the results are shown to emphasize the limited applicability 
of this approximation.
The technical detail of our resonant BCS plus QRPA calculation
is explained in the previous report \cite{Ya04}.
The box size of $R_{box}=20$ fm is used 
only for the resonant BCS approximation.
The results have the smooth dependence on $R_{box}$ 
up to around 20 fm, and they are consistent with
our preliminary results obtained by the resonant BCS plus QRPA 
calculation with the constant pairing gap approximation \cite{Ya04}. 
However we couldn't obtain the regular solution 
in $^{84, 86}$Ni with the larger $R_{box}$. 
In the present study,  as an improvement,
the density-dependent pairing interaction of 
Eq.(\ref{EQ-DDPI}) is used.
Because the cut-off energies of the particle model space 
are different in the resonant BCS calculation and the QRPA calculation,
the different pairing strengths must be used for them.
The pairing strength is changed for each nucleus 
to reproduce the average neutron pairing gap
obtained by HFB calculation.
On the other hand, for the QRPA residual interaction, 
exactly the same interaction used in the HFB plus QRPA calculation
is adopted.

From $^{80}$Ni to $^{82}$Ni, the results of 
the HFB plus QRPA  and  the resonant BCS plus QRPA  
have a similar behavior; namely
the energies decrease and the $B(E2)$ values increase.
On the other hand, 
as approaching the neutron drip line, the $B(E2)$ values 
obtained by the resonant BCS plus QRPA calculation
are much smaller than those in the HFB plus QRPA calculation. 
The main qualitative difference between these calculations
is the realization of the broad localization of 
the two-quasiparticle states associated with
the neutron low-lying $s_{1/2}$, $d_{3/2}$ and $d_{5/2}$ 
resonant states in HFB.

The results for $^{84, 86}$Ni obtained by RPA calculation 
are also shown in Fig.\ref{FIG_Ni}.
The $B(E2)$ values in RPA is much smaller than those
in the resonant BCS plus QRPA calculation, 
because the contributions of the hole-hole channel can
increase the collectivity of these excitations.
As we have discussed in Subsec.\ref{SUBSEC-TQP},
the lowest discrete solutions in RPA are 
single-particle like excitations.
The excitation energies are almost the same with the specific
\opoh energies ($2d_{5/2}$ $\to$ $3s_{1/2}$ in $^{84}$Ni, 
and $3s_{1/2}$ $\to$ resonant $d_{3/2}$ in $^{86}$Ni).
In Fig.\ref{FIG_RPA} the isoscalar and charge quadrupole strengths 
in $^{86}$Ni obtained by RPA calculation, 
and the corresponding unperturbed strengths are shown.
The neutron \opoh excitations; 
(a) $3s_{1/2}$ $\to$ resonant $d_{3/2}$,
(b) $2d_{5/2}$ $\to$ resonant $d_{3/2}$, and
(c) $2d_{5/2}$ $\to$ resonant $g_{7/2}$,
are indicated.
The excitation,
(x) $3s_{1/2}$ $\to$ (discretized) non-resonant continuum $d_{5/2}$, 
is also pointed out. 

The excitations to the resonant states, (a), (b) and (c),
can generate some collectivity among the localized components
(see Figs.\ref{FIG_Fph} and \ref{FIG_Rph}).
However the $B(E2)$ values of the first 2$^+$ states in RPA change
from 234 $e^2$fm$^{4}$ to 173 $e^2$fm$^{4}$ in $^{84}$Ni, 
from 83.0 $e^2$fm$^{4}$ to 74.5 $e^2$fm$^{4}$ in $^{86}$Ni
as increasing the box size from 20 fm to 30 fm.
The HF calculation with small $R_{box}$ leads to
the artificial localization of the spatially extended 
wave functions. Accordingly the RPA calculation
overestimates the correlations among these \opoh states.
To obtain fully convergent solutions in RPA, 
much larger $R_{box}$ or continuum RPA calculation
is required.
The excitations to non-resonant continuum states like
(x) can't produce the collectivity.
The small peak related to (x) in the charge quadrupole strength 
is an artifact of the box boundary condition, and the strength
decreases with larger $R_{box}$.



\section{Conclusion}  \label{SEC-CONC}

We have investigated the role of low-$\ell$ neutrons and 
pairing correlations for low-frequency vibrational excitations 
in neutron drip line nuclei.
We have clarified the change of the spatial structure
of quasiparticle wave functions induced 
by self-consistent pairing correlation;
the pairing anti-halo effect in the lower component
and the broadening effect in the upper component.
We have found that the broad localization of 
two-quasiparticle states among low-$\ell$ neutrons is realized
in consequence of the structure change in 
the quasiparticle wave functions.
The broad localization can cause novel features of
low-frequency vibrational excitations in nuclei 
close to the neutron drip line;
as demonstrated by HFB plus QRPA calculation 
for the first 2$^+$ states in neutron rich Ni isotopes, 
it brings about the spatially extended coherence 
and the large transition strength.


\begin{acknowledgments}
I acknowledge 
Professor K. Matsuyanagi for valuable discussions.
I thank Professor I. Hamamoto, Professor H. Sagawa,
and Professor M. Matsuo for useful comments.
I also thank discussions with the members of the
Japan-U.S. Cooperative Science Program
"Mean-Field Approach to Collective Excitations in Unstable
Medium-Mass and Heavy Nuclei".
I am grateful for the financial assistance from 
the Special Postdoctoral Researcher Program of RIKEN. 
Numerical computation in this work was carried out at the 
Yukawa Institute Computer Facility.
\end{acknowledgments}


%
%

\end{document}